\documentclass[nofootinbib,a4paper,aps,prd,10pt,superscriptaddress,showkeys,twocolumn]{revtex4}

\usepackage{graphicx}
\usepackage{amsfonts}
\usepackage{amssymb}
\usepackage{amsmath}
\usepackage{hyperref}
\usepackage{natbib}
\usepackage{float}
\usepackage{dcolumn}
\usepackage{bm}
\usepackage{latexsym,color}

\def\prl{Phys. Rev. Lett.}

\def\prd{Phys. Rev. D}

\def\mnras{Mon. Not. Roy. Astr. Soc.}
\def\apj{Astrophys. J.}
\def\apjl{Astrophys. J. Lett.}

\newcommand{\bea}{\begin{eqnarray}}
\newcommand{\eea}{\end{eqnarray}}
\newcommand{\be}{\begin{equation}}
\newcommand{\ee}{\end{equation}}

\begin{document}

\title{Luminosity of accretion disks around rotating regular black holes}

\author{Kuantay~\surname{Boshkayev}}
\email[]{kuantay@mail.ru}
\affiliation{National Nanotechnology Laboratory of Open Type,  Almaty 050040, Kazakhstan.}
\affiliation{Al-Farabi Kazakh National University, Al-Farabi av. 71, 050040 Almaty, Kazakhstan.}
\affiliation{International Information Technology University, Manas st. 34/1, 050040 Almaty, Kazakhstan.}

\author{Talgar~\surname{Konysbayev}}
\email[] {talgar\_777@mail.ru}
\affiliation{National Nanotechnology Laboratory of Open Type,  Almaty 050040, Kazakhstan.}
\affiliation{Al-Farabi Kazakh National University, Al-Farabi av. 71, 050040 Almaty, Kazakhstan.}

\author{Yergali~\surname{Kurmanov}}
\email[]{kurmanov.yergali@kaznu.kz}
\affiliation{National Nanotechnology Laboratory of Open Type,  Almaty 050040, Kazakhstan.}
\affiliation{Al-Farabi Kazakh National University, Al-Farabi av. 71, 050040 Almaty, Kazakhstan.}

\author{Orlando~\surname{Luongo}}
\email[]{orlando.luongo@unicam.it}
\affiliation{Al-Farabi Kazakh National University, Al-Farabi av. 71, 050040 Almaty, Kazakhstan.}
\affiliation{Universit\`a di Camerino, Via Madonna delle Carceri 9, 62032 Camerino, Italy.}
\affiliation{SUNY Polytechnic Institute, 13502 Utica, New York, USA.}
\affiliation{Istituto Nazionale di Fisica Nucleare, Sezione di Perugia, 06123, Perugia,  Italy.}
\affiliation{INAF - Osservatorio Astronomico di Brera, Milano, Italy.}

\author{Marco~\surname{Muccino}}
\email[] {marco.muccino@lnf.infn.it}
\affiliation{Al-Farabi Kazakh National University, Al-Farabi av. 71, 050040 Almaty, Kazakhstan.}

\author{Aliya~\surname{Taukenova}}
\email[] {aliya.tauken@gmail.com}
\affiliation{National Nanotechnology Laboratory of Open Type,  Almaty 050040, Kazakhstan.}
\affiliation{Al-Farabi Kazakh National University, Al-Farabi av. 71, 050040 Almaty, Kazakhstan.}

\author{Ainur~\surname{Urazalina}}
\email[] {y.a.a.707@mail.ru}
\affiliation{National Nanotechnology Laboratory of Open Type,  Almaty 050040, Kazakhstan.}
\affiliation{Al-Farabi Kazakh National University, Al-Farabi av. 71, 050040 Almaty, Kazakhstan.}

\date{\today}

\begin{abstract}
We consider thin accretion disks in the field of a class of rotating regular black holes. For this purpose, we obtain the radius of the innermost stable circular orbit, $r_{ISCO}$ and efficiency of accretion disk in converting matter into radiation $\eta$ with the aim of modeling the disk's emission spectrum. We consider a simple model for the disk's radiative flux, differential and spectral luminosity and compare the results with those expected from accretion disks around Kerr black holes. As a remarkable result, from our computations we find that both the luminosity of the accretion disk and the efficiency are larger in the geometry of rotating regular black holes for fixed and small values of the spin parameter $j$ with respect to those predicted with the Kerr metric for a black hole of the same mass. These results may have interesting implications for astrophysical black holes.
\end{abstract}



\maketitle

\section{Introduction}

The existence of black holes (BHs) comes as a direct byproduct of the equations of general relativity and today plays a pivotal role in astrophysics. The presence of astrophysical BH candidates\footnote{Even though BHs candidates can be considered as consolidated astrophysical objects, some open issues related to their nature and observational constraints are still open \cite{Cardoso:2019rvt, Vagnozzi:2019apd, Perlick:2021aok}.} is directly supported, for instance, by the discovery of gravitational waves originating from binary mergers  \cite{Abbott,Abbott2018,Abbott2016,Abbott2017,AbbottPhysRevLett,Abbott2017ApJ} and by the recent imaging of the ``shadow'' of the compact objects residing at the centers of the galaxy M87 and the Milky Way \cite{EventHorizonTelescope:2019dse, EventHorizonTelescope:2019ggy, EventHorizonTelescope:2022wkp}.
For example, from the image of a BH shadow one cannot exclude alternative models, including BH mimickers \cite{Abdujabbarov:2016hnw, Li:2022eue, Abdikamalov:2019ztb, Olivares:2018abq} and naked singularities \cite{Shaikh:2018lcc, Bambhaniya:2021ybs, Guo:2020tgv, Dey:2020bgo, Joshi:2020tlq}, as due to the degeneracy among competing models within the observational error bars.

In addition, when the curvature of space-time approaches and exceeds the Planck value, the notion of a classical space-time can be misleading and the general consensus is that singularities must be resolved within a theory of quantum-gravity. This raises the question of whether astrophysical BHs are well-described by the BHs solutions known from general relativity or whether the models of BH mimickers and/or non-singular BHs may be better suited.

In this respect, regular black holes (RBHs) are arguably the most interesting objects that elude the conditions of the singularity theorems while potentially retaining the properties of astrophysical BH candidates. Static and rotating RBHs have been derived in a wide variety of scenarios \cite{Bronnikov1979,Gonzalez-Diaz,Poisson,Dymnikova,Bronnikov:2006fu, Balart:2014cga, Fan:2016hvf, Toshmatov:2018cks, Toshmatov:2017zpr, Carballo-Rubio:2018pmi, Gibbons:1985ac, Bambi2013PhLB, Borde1994PhRvD,Barrabes,Bogojevic,Cabo,Hayward201,Jusufi2020,Ghosh2015EPJC,ToshmatovPhysRevD,Mustapha2014PhRvD} and some of these models can be  dynamically obtained from gravitational collapse \cite{Malafarina:2022oka, Carballo-Rubio:2023mvr}. Therefore it is not surprising that the study of their observational properties is currently an active research direction \cite{Flachi:2012nv, Stuchlik:2014qja, Toshmatov:2015wga, Toshmatov:2019gxg, Toshmatov:2021fgm, Stuchlik:2019uvf, Singh2022,KumarApj,Ghosh2022}.
In this respect, recent efforts showed how the Bardeen \cite{Bardeen} and Hayward \cite{Hayward} RBHs are expected to modify the spectral luminosity of the accretion disk with respect to that of the Kerr and Schwarzschild BHs \cite{Akbarieh:2023kjv}.

Motivated by the above considerations on RBHs, we here consider the possibility of distinguishing \emph{rotating} RBHs from Kerr BHs by means of observations of the emission spectrum of the accretion disk. To do so, we adopt the approach pioneered by Novikov, Page and Thorne, see e.g.  \cite{novikov1973, page1974}, to examine  RBHs surrounded by accretion disk. In this respect, our analysis revolves around a basic conceptual model of the accretion disk, incorporating a set of simplifications that enable us to effectively predict all spectral and thermodynamic variables. Under the above recipe, we assume a thin accretion disk, described by  $\delta$-like distribution of matter located on the equatorial plane. Following Ref. \cite{Bambi:2014koa}, we assume the mass of the disk to be negligible with respect to that of the central object. This enables us to consider geodesic motion in the background geometry of the central object, making the assumption that the particles within the disk lack charge and follow circular orbits. Consequently, we disregard the effects of dynamical friction, which would otherwise cause the particles to spiral inward. Nevertheless, given a sufficiently short timescale, this effect can be neglected and so we further adopt a straightforward emission model for the disk particles, resulting in a \emph{black body spectrum}.

By utilizing the aforementioned assumptions, we calculate the innermost stable circular orbit (ISCO), the flux, the differential luminosity, the spectral luminosity, and the efficiency of mass-to-energy conversion within the accretion disk. Finally, we compare our findings with their counterparts in the Kerr geometry.

The paper is organized as follows. In Sect. \ref{sez2}, we describe the main features of the class of rotating RBHs considered, briefly reviewing  the  formalism for test particle motion in accretion disks. In Sect. \ref{sez3}, we review the thin accretion disk formalism from the Novikov-Thorne and Page-Thorne models, applying it to the rotating RBH solution under consideration. We thus discuss and compare our findings with the corresponding results in the Kerr spacetime. Finally, in section \ref{sez5} we discuss the potential implications of our results for the observation of astrophysical BHs. Throughout the paper we use natural units setting $G=c=1$.


\section{Test particle motion around regular black holes}\label{sez2}

In this section, we examine a group of RBH solutions in a general form, which incorporates the cosmological constant  \cite{2014PhLB..734...44N} and derive the orbital quantities of neutral test particles on circular orbits in the equatorial plane.

The general solution considered here is built from a mass function of the type
\begin{equation}
\label{massfunc}
m(r) = M\,\left[1 + \left(\frac{r_0}{r}\right)^q \right]^{-\frac{p}{q}}\,,
\end{equation}
which in the static case without cosmological constant guarantees an asymptotically flat spacetime for positive $p$ and $q$. It has been shown that RBHs static solutions of this kind can be obtained from gravity coupled to a theory of non-linear electrodynamics \cite{Ayon-Beato:1998hmi, Bronnikov:2000yz, Bronnikov:2022ofk}.
Here $M$ and $r_0$ are the mass and length parameters, respectively. The well-known Bardeen \cite{Bardeen,Ansoldi} and Hayward \cite{Hayward}  BHs correspond to the choices $p=3$, $q=2$ and $p=q=3$, respectively and the choice $p\geq 3$ ensures that the geometry is regular at the center for the static RBHs.

The most general line element including rotation and the cosmological constant obtained from the above mass function
is written in Boyer-Lindquist coordinates as
\begin{align}
ds^{2} &=-\frac{1}{\Sigma}\left(\Delta_{r}-\Delta_{\theta}a^{2}\sin^2\theta\right)dt^{2} + \frac{\Sigma}{\Delta_{r}}dr^{2} + \frac{\Sigma}{\Delta_{\theta}}d\theta^{2}\nonumber \\
& + \frac{1}{\Xi^{2}\Sigma}\left[(r^{2}+a^{2})^{2}\Delta_{\theta}-\Delta_{r}a^{2}\sin^2\theta\right]\sin^2\theta d\phi^{2}\nonumber \\
& - \frac{2a}{\Xi\Sigma}\left[(r^{2}+a^{2})\Delta_{\theta}-\Delta_{r}\right]\sin^2\theta dtd\phi ,
\label{Metrica_Boyer-Lindquist}
\end{align}
 where
\begin{subequations}
    \begin{align}
&\Delta_{\theta}=1+\frac{\Lambda}{3}a^{2}\cos^2\theta, \ \Sigma=r^{2}+a^{2}\cos^2\theta, \label{definitions} \\
&\Delta_{r}=\left(r^{2}+a^{2}\right)\left(1-\frac{\Lambda}{3}r^{2}\right)-2rm, \ \Xi=1+\frac{\Lambda}{3}a^{2}, \label{x,Deltar} \
    \end{align}
\end{subequations}
where $a$ is the Kerr parameter related to the angular momentum of the source.
For vanishing $\Lambda$ and $r_0$ the metric Eq. \eqref{Metrica_Boyer-Lindquist} reduces to the Kerr solution, that will be  considered as a reference model to compare our RBH expectations.

\subsection{Conditions on the metric}

In the following, for the sake of simplicity, we consider the Hayward-like rotating RBH, by setting $p=q=3$, without cosmological constant, i.e. $\Lambda=0$. The RBH obtained for $p=q=3$ is also preferable because in the non-rotating case it describes a simple model of a BH in general relativity coupled to a simple theory of non-linear electrodynamics which can be obtained dynamically from a collapse scenario similarly to the classical Oppenheimer-Snyder model \cite{Malafarina:2022oka}.

\begin{figure*}[ht]
\begin{minipage}{0.49\linewidth}
\center{\includegraphics[width=0.99\linewidth]{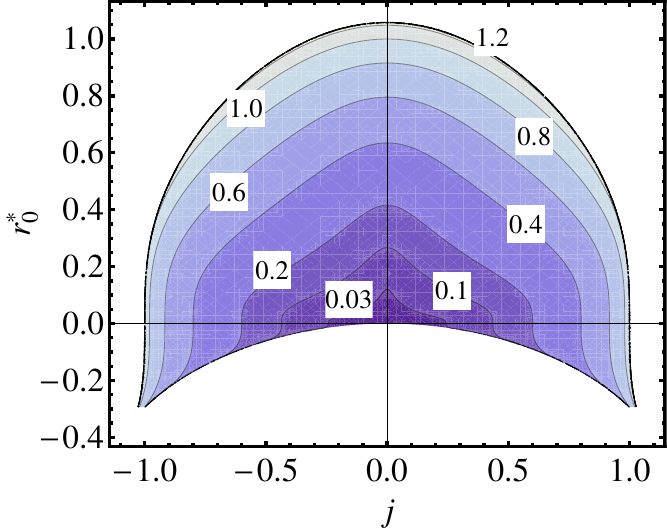}\\ }
\end{minipage}
\hfill
\begin{minipage}{0.50\linewidth}
\center{\includegraphics[width=0.95\linewidth]{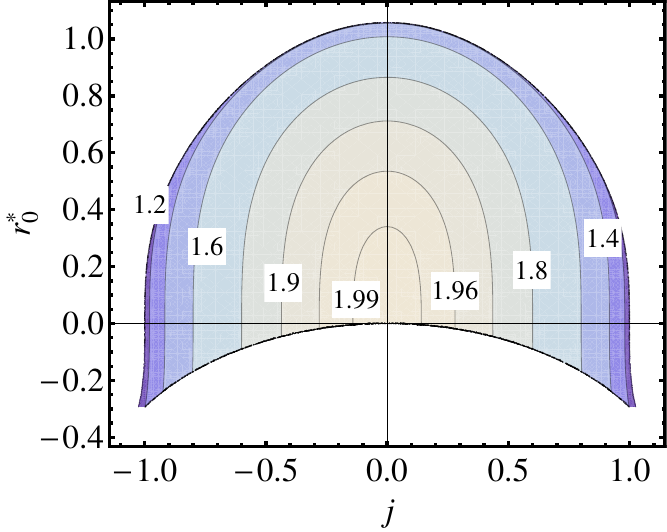}}\\
\end{minipage}
\caption{The figure consists of contour plots depicting the locations of the inner (left panel) and outer (right panel) horizons. These locations are presented as functions of the deformation parameter $r_0^{*}=r_0/M$ and the spin parameter $j=a/M$ for the rotating Hayward black hole. It is important to note that the shaded region represents the parameter values for which the horizons exist. The numerical values displayed on the contours indicate the corresponding horizon radius $r_H/M$. It should be observed that when $r_0^{*}\neq 0$, the values of the horizon radii satisfying the extremality condition may exceed 1.}
\label{fig:horizon}
\end{figure*}

Similarly to the Kerr solution, also the rotating solution, under the consideration, has an inner and an outer horizons, useful to characterize the accretion disk properties. Specifically, the horizons of rotating Hayward BHs is calculated using the standard condition $1/g_{rr}=0$. Its solution  contains five roots, where only two of them are physical. The remaining three roots are complex and thus non-physical. The largest out of the two physical roots corresponds to the outer horizon and the smallest corresponds to the inner horizon in analogy with the Kerr metric. It should be noted that unlike the static case where $r_0$ can be only positive, for a rotating case one can in principle allow for $r_0$ to be negative. In the case of a RBH obtained from a theory of non-linear electrodynamics a negative $r_0$ could then be interpreted as related to the sign of the magnetic charge. The values of the inner and outer horizons depending on the parameters $r_0^*=r_0/M$ and $j=a/M$ are plotted in Fig.~\ref{fig:horizon}.

It should be noted that for every given value of $r_0^*$, there exists a range of $j$ values within which both horizons exist. However, there is a special extremal case where the two horizons coincide, resulting in a boundary beyond which the spacetime becomes horizonless. Furthermore, it is worth noting that in the Kerr case, where $r_0^*=0$, both the inner and outer horizons approach the limit $r_H/M\rightarrow 1$ in the extremal scenario as $j\rightarrow1$. However, when $r_0^* \neq 0$, the limiting value of $r_H/M$ at which the two horizons coincide is larger than unity.

\subsection{Circular orbits for rotating regular black holes}

For circular orbits in the equatorial plane we set $\theta=\pi/2$, with  $\dot{r}=\dot{\theta}=\ddot{r}=0$. The geodesic equation leads to the following relation for the angular velocity of test particles
\begin{equation}
\Omega=\frac{aM\left(r^{3}-2r^{3}_{0}\right)-\sqrt{M\left(r^{3}-2r^{3}_{0}\right)}A}{a^{2}M\left(r^{3}- 2r^{3}_{0}\right)- A^{2}},
\label{11}
\end{equation}
where $A=A(r)=r^{3}+r^{3}_{0}$ and $a>0$ ($a<0$) corresponds to co-rotating (counter-rotating) particles with respect to the direction of rotation of the BH, respectively.

The  energy per unit mass ${E}$ and the angular momentum per unit mass ${L}$ of test particles moving in circular orbits are then given by
\begin{subequations}
\begin{align}
{{E}}&=\frac{C_-+aB\Omega}{\sqrt{A\left(C_-+2aB\Omega-\left(Ar^{2}+a^{2}C_+\right)\Omega^{2}\right)}},
\label{12}\\
{{L}}&=\frac{\left(Ar^{2}+a^{2}C_+\right)\Omega-2aB}{\sqrt{A\left(C_-+2aB\Omega-\left(Ar^{2}+a^{2}C_+\right)\Omega^{2}\right)}},
\label{13}
\end{align}
\end{subequations}
where
\begin{eqnarray}
 B=2Mr^{2},\quad
 C_{\pm}=A\pm B.
\end{eqnarray}

In order to  find $r_{ISCO}$, we will simply use $dL/dr=dE/dr=0$ condition, which is equivalent to $d^2U_{eff}/dr^2=0$, where $U_{eff}$ is the effective potential of test particles in circular orbits.

\begin{figure*}[ht]
\begin{minipage}{0.49\linewidth}
\center{\includegraphics[width=0.97\linewidth]{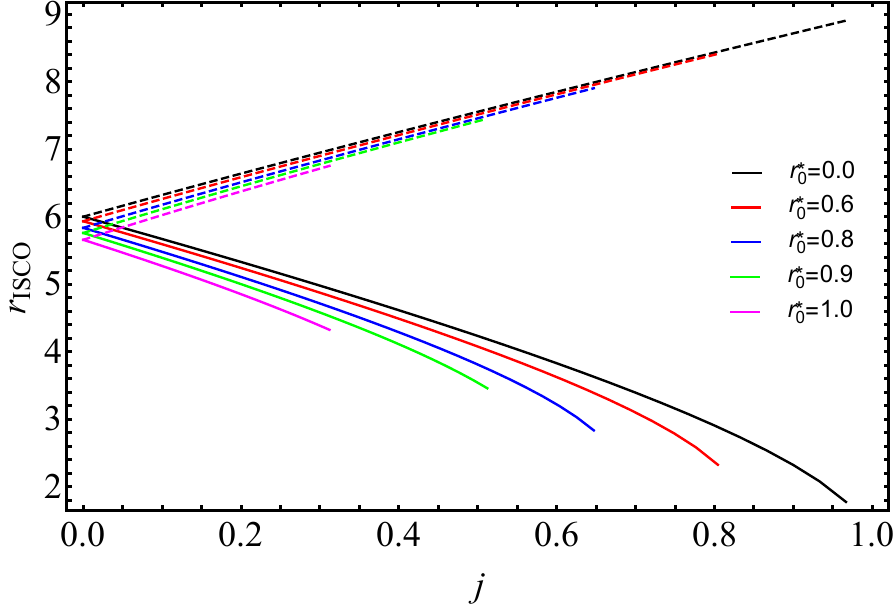}\\ }
\end{minipage}
\hfill
\begin{minipage}{0.50\linewidth}
\center{\includegraphics[width=0.97\linewidth]{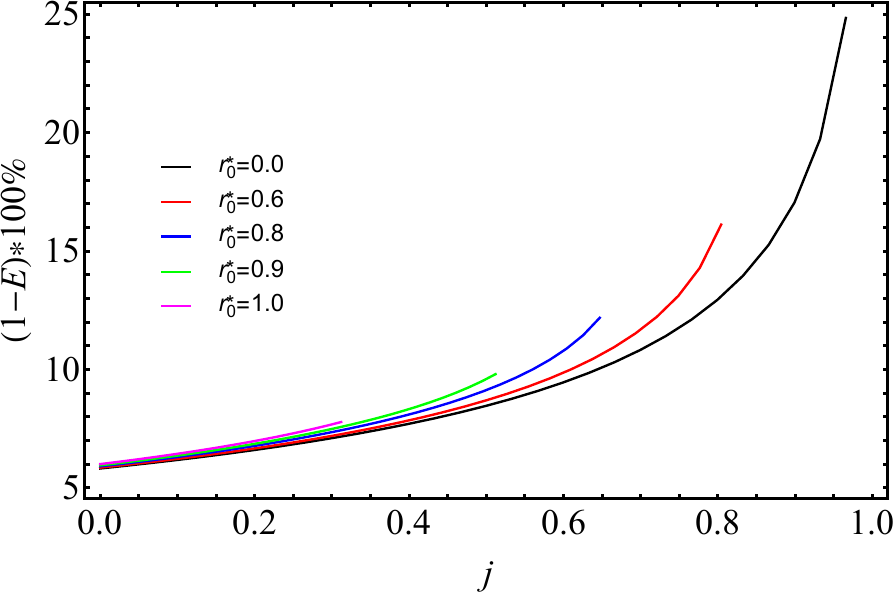}\\ }
\end{minipage}
\caption{Left panel: The ISCO radii as a function of the angular momentum $j$ for the rotating RBH with different values of $r_0^*$. Right panel: The radiative efficiency $\eta$ of rotating RBHs as a function of the angular momentum $j$ for different values of the parameter $r_0^*$. The case $r_0^*=0$ corresponds to the Kerr BH. The curves end at the values of $j$ where the RBH horizon becomes extremal.}
\label{fig:riscoj,Effiiencyj}
\end{figure*}

\begin{figure*}[ht]
\begin{minipage}{0.49\linewidth}
\center{\includegraphics[width=0.99\linewidth]{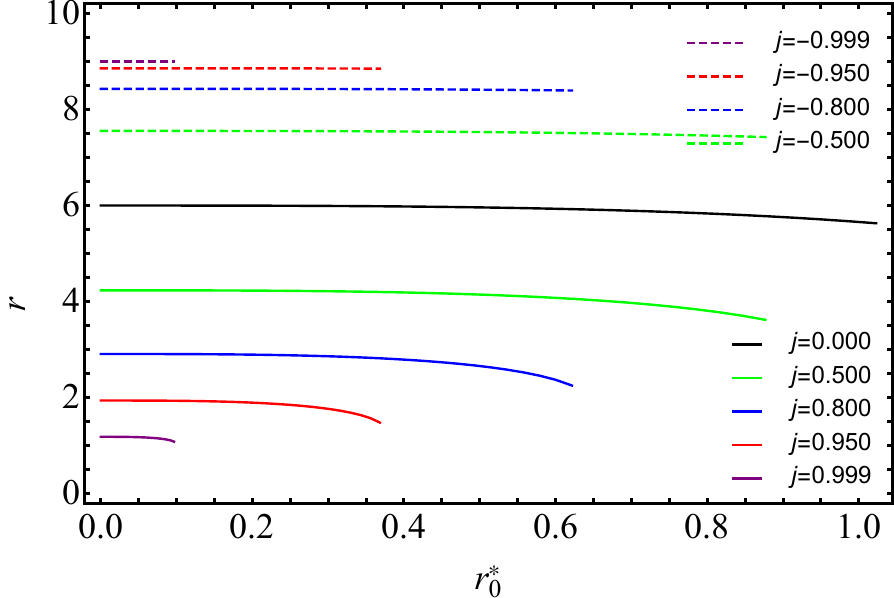}\\ }
\end{minipage}
\hfill
\begin{minipage}{0.50\linewidth}
\center{\includegraphics[width=0.95\linewidth]{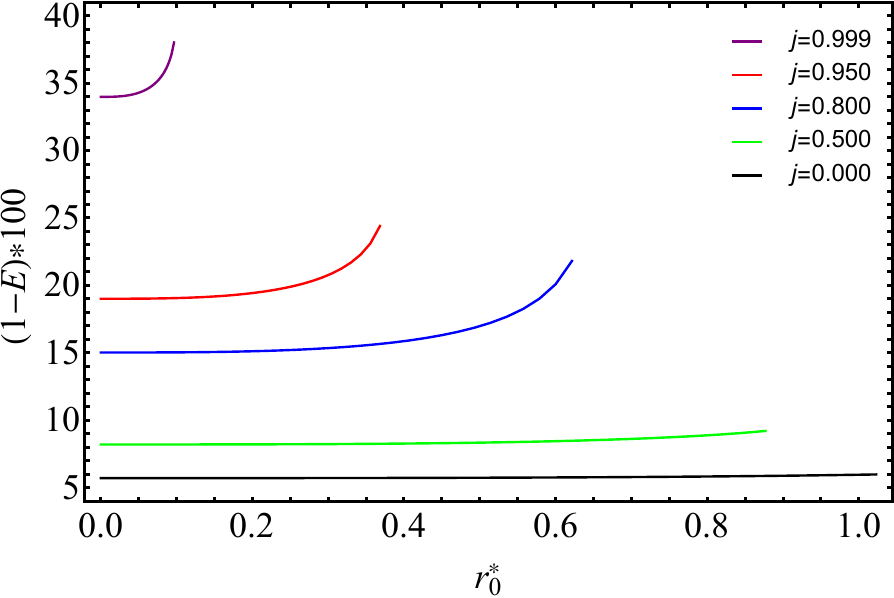}\\ }
\end{minipage}
\caption{Left panel: The ISCO radii as a function of the parameter $r_{0}^{*}$ for the rotating RBH (the blue, red, green and purple curves) and Hayward metric (the black solid curve) for different values of the angular momentum $j$. Right panel: Radiative efficiency $\eta$ as a function of the parameter $r_{0}^{*}$ (the blue, red, green and purple curves) and Hayward metric (the black solid curve) for different values of the angular momentum $j$. The curves end at the values of $j$ where the RBH horizon becomes extremal.}
\label{fig:riscor0,Effiiencyr0}
\end{figure*}

\subsection{Comparing with the Kerr solution}

With the aim of comparing accretion disks surrounding rotating RBHs with the Kerr mettric, we now briefly review the relevant quantities for particles on circular orbits in the Kerr geometry, whose line element is \cite{1963PhRvL..11..237K}
\begin{eqnarray}
ds^2&=&-\left(1-\frac{2Mr}{\Sigma}\right)dt^{2}+\frac{\Sigma}{\Delta}dr^{2}+\Sigma d\theta^{2}\nonumber\\
&+&\left(r^{2}+a^{2}+\frac{2Mra^{2}}{\Sigma}\sin^2\theta\right)\sin^2\theta d\phi^{2}\nonumber\\
&-&\frac{4Mra}{\Sigma}\sin^{2}\theta d\phi dt,
\label{eq:metricKerr}
\end{eqnarray}
where $\Sigma=r^{2}+a^{2}\cos^{2}\theta$ and $\Delta=r^{2}-2Mr+a^{2}$. The total gravitational mass of the source is given by $M$, its dimensionless angular momentum, namely  the spin parameter, is  $j=a/M$ and so one conclude that the Kerr metric is fully-characterized by two parameters only. The Schwarzschild metric is recovered as $a=0$.

For the Kerr solution the angular velocity, angular momentum and energy of the particle moving along the circular orbits are

\begin{subequations}
\begin{align}
 \label{eq:omegaKerr}
\Omega^{2}&= \frac{M}{r^{3}\pm 2a r^{2}\sqrt{M/r} +a^{2}M}, \\
 \label{eq:energyKerr}
E^{2}&=\frac{\left(\sqrt{r}\left(r-2M \right)\pm a \sqrt{M}\right)^{2}}{r^{2}\left(r\pm 2a\sqrt{M/r}-3M\right)}, \\
 \label{eq:angmomKerr}
L^{2}&= \frac{M\left(r^{2}\mp 2a\sqrt{M/r}+a^{2}\right)^{2}}{r^{2}\left(r\pm 2a\sqrt{M/r}-3M\right)},
\end{align}
\end{subequations}
where the $\pm$ signs correspond to co-rotating and
counter-rotating particles with respect to the direction of rotation of the BH, respectively \cite{1972ApJ...178..347B}.

The radius of the ISCO for the Kerr metric is given by
\begin{equation}
 \label{eq:riscoKerr}
\frac{r_{ISCO}^{\pm}}{M}=3+Z_2\pm \sqrt{(3-Z_1)(3+Z_1+2Z_2)},
\end{equation}
with
\begin{subequations}
\begin{align}
Z_1&\equiv 1+\left(1-j^2\right)^{\frac{1}{3}}\left(\left(1+j\right)^{\frac{1}{3}}+\left(1-j\right)^{\frac{1}{3}}\right),\label{eq:Z1}\\
Z_2&\equiv \left(3 j^2+Z^2_1\right)^{\frac{1}{2}}.\label{eq:Z2}
\end{align}
\end{subequations}
where the $\pm$ signs correspond to counter-rotating  and co-rotating particles.

At this point, it is beneficial to introduce the BH efficiency, $\eta$,  in converting matter into radiation. Unlike the location of the ISCO, this quantity is \emph{coordinate-independent} and can be measured, at least in principle. It is given by
\be
\eta=[1-E(r_{ISCO})]\times100\%\,,
\ee
playing a prominent role in the physics of accretion disks.

Thus, to carry out the numerical analysis and present results more explicitly, it is useful to introduce dimensionless quantities defined as\footnote{The behavior of $\Omega^*$, $L^*$ and $E^*$ for Kerr and the static RBH is illustrated in the appendix. As expected departures from Schwarzschild are more pronounced at small radii.} $\Omega^*(r)=M\Omega(r)$, $L^*(r)=L(r)/M$, $E^*(r)=E(r)$ and $r_0^*=r/M$.

In Fig.~\ref{fig:riscoj,Effiiencyj}, we display the ISCO radius, $r_{ISCO}$, (left panel) and the efficiency, $\eta$, (right panel) of accretion disks as functions of the spin parameter $j$, selecting  different values of $r_0^*$. In Fig.~\ref{fig:riscor0,Effiiencyr0}, we display the ISCO radius, $r_{ISCO}$, (left panel) and the efficiency, $\eta$, (right panel) of accretion disks as functions of the parameter $r_{0}^{*}$ for different values of $j$. As one may notice, that in order to get smaller $r_{ISCO}$ and correspondingly larger $\eta$, the spin parameter $j$ must increase and the parameter $r_0^*$ must decrease. Eventually, one can get the smallest $r_{ISCO}$ and largest $\eta$, only when $r_0^*=0$ i.e. $j=1$, which corresponds to the extreme Kerr black hole. However, for larger values of $r_0^*$ and smaller values of $j$, the rotating RBHs can have smaller  $r_{ISCO}$ and correspondingly larger $\eta$ with respect to the Kerr black holes.

Recalling that the Kerr spacetime is obtained for $r_0^*=0$ and the static Hayward BH is obtained for $j=0$, we notice that the curves end at the values of $j$ and $r_0^*$ for which the RBH becomes extreme, consistently with what shown in Fig.~\ref{fig:horizon}.

\section{Spectra of thin accretion disks}\label{sez3}

To explore the luminosity and spectral properties of the accretion disk, surrounding RBHs, we involve the simplest approach developed in \cite{novikov1973} and \cite{page1974}, where the radiative flux\footnote{The radiative flux is commonly the energy emitted per unit area per unit time from the accretion disk. }, $\mathcal{F}$, yields
\begin{equation}
 \label{eq:flux}
\mathcal{F}(r)=-\frac{\mathcal K}{4\pi \sqrt{-g}} \frac{\Omega_{,r}}{\left(E-\Omega L\right)^2 }\int^r_{r_{ISCO}} \left(E-\Omega L\right) L_{,\tilde{r}}d\tilde{r},
\end{equation}
for a constant mass accretion rate, defined by $\mathcal K\equiv\dot{{\rm m}}$, consequently computing the flux per unit accretion rate $\mathcal{F}/\dot{{\rm m}}$ in lieu of the flux itself.

Spacetime geometry is clearly involved through the determinant of the metric of the three-dimensional sub-space   ($t,r,\varphi$), i.e. $\sqrt{-g}=\sqrt{-g_{rr}(g_{tt}g_{\varphi\varphi}-g_{t\varphi}^2)}$ \cite{2012ApJ...761..174B}, where for the Kerr and rotating Hayward BH $\sqrt{-g}=r$ when $\theta=\pi/2$.
Thus, invoking thermodynamic equilibrium, the corresponding radiation appears emitted by virtue of a black body. Hence, radiation is modelled as isotropic, providing the Stefan-Boltzmann law for the emitted temperature
\begin{equation}
 \label{eq:temperature}
\mathcal{F}(r)=\sigma\,T^{*4},
\end{equation}
where $\sigma$ is the usual Stefan--Boltzmann constant and $\mathcal F(r)$ the radiative flux, function of the radial distance, whereas $T^*$ the intrinsic temperature.

\subsection{Observable quantities from accretion disks}

Remarkably, it appears crucial to emphasize that the radiative flux $\mathcal{F}$ is not directly observable, as it is a quantity measured in the rest frame of the accretion disk.

From an observational perspective, the differential luminosity $\mathcal{L}_{\infty}$ appears more practical than radiative fluxes that, conversely, cannot be measured. Following Refs. \cite{novikov1973, page1974}, we write the differential luminosity as
\begin{equation}
 \label{eq:difflum}
\frac{d\mathcal{L}_{\infty}}{d\ln{r}}=4\pi r \sqrt{-g}E \mathcal{F}(r),
\end{equation}
that clearly describes the radiation emitted by the accretion disk at a given distance.

From the differential luminosity, one can determine spectra and frequencies, i.e., the direct measurable quantities.

Defining the spectral luminosity distribution, one computes it as observed at infinity, with the recipe of providing a  black body, namely  \cite{2020MNRAS.496.1115B}
\begin{equation}
 \label{eq:speclum}
\nu \mathcal{L}_{\nu,\infty}=\frac{60}{\pi^3}\int^{\infty}_{r_{ISCO}}\frac{\sqrt{-g }E}{M_T^2}\frac{(u^t y)^4}{\exp\left[u^t y/\mathcal{F}^{*1/4}\right]-1}dr,
\end{equation}
where $\mathcal{L}_{\nu,\infty}$ points out that observations are provided far from the solution. In the above relation, $y=h\nu/kT^*$, $h$ is the Planck constant, $\nu$ is  the emitted radiation frequency and $k$ is the Boltzmann constant.

Moreover, we have
\begin{equation}
 \label{eq:sample7}
u^t(r)=\frac{1}{\sqrt{-g_{tt}-2\Omega g_{t\varphi}-\Omega^2 g_{\varphi \varphi}}},
\end{equation}
where $u^t$ is the contra-variant zero (time) component of the four-velocity.

The dimensionless argument implies normalized flux with respect to the total mass $M$. For our computations, the quantity $\mathcal{F}^{*}(r) = M^2 \mathcal{F}(r)$ yields a well-posed representation of the flux that we will display in our numerical findings.

\begin{figure*}[ht]
\begin{minipage}{0.49\linewidth}
\center{\includegraphics[width=0.98\linewidth]{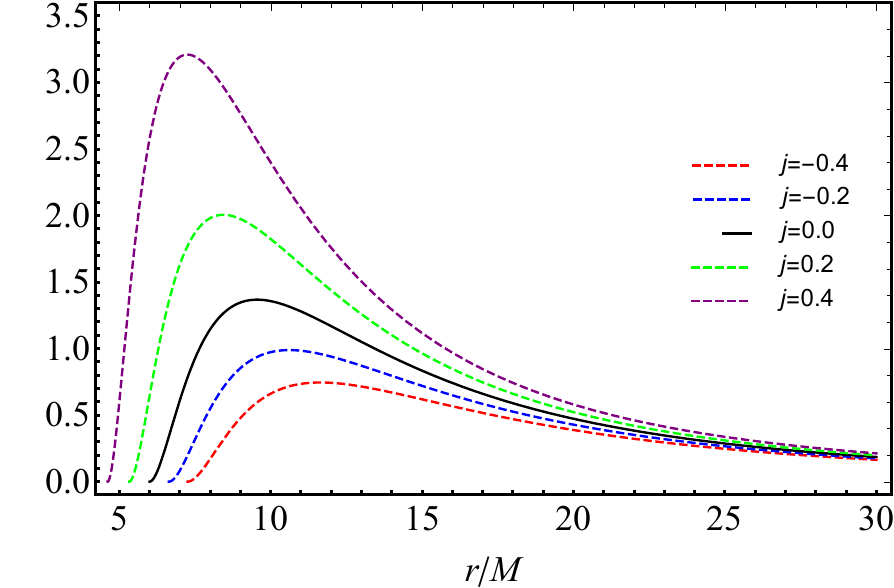}\\ }
\end{minipage}
\hfill
\begin{minipage}{0.50\linewidth}
\center{\includegraphics[width=0.98\linewidth]{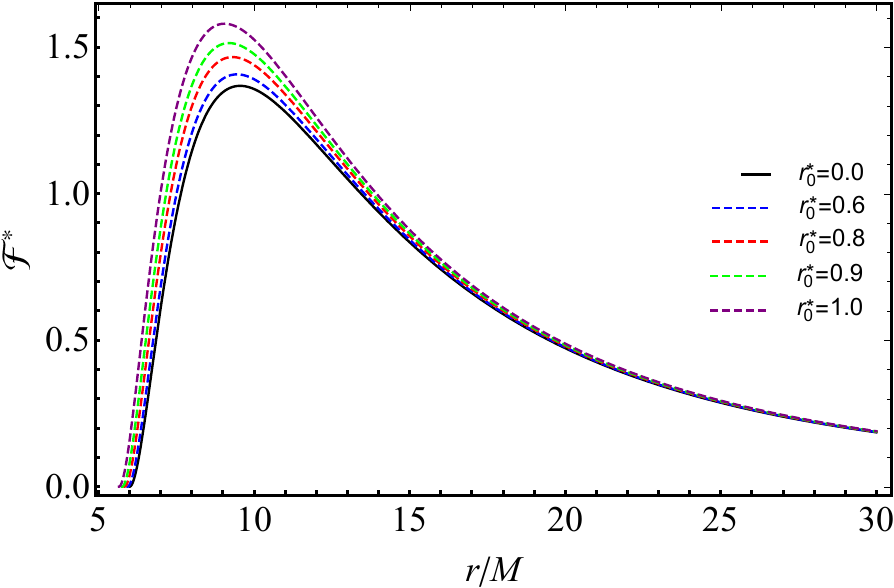}\\ }
\end{minipage}
\caption{Left panel: Radiative flux $\mathcal{F}^*$ multiplied by $10^{5}$ of the accretion disk versus normalized radial distance $r/M$ for rotating RBHs with $r_{0}^{*}=0$. Right panel: Radiative flux $\mathcal{F}^*$ multiplied by $10^{5}$ of the accretion disk versus normalized radial distance $r/M$ for RBHs with $j=0$.}
\label{fig:FluxKr0}
\end{figure*}

\begin{figure*}[ht]
\begin{minipage}{0.49\linewidth}
\center{\includegraphics[width=0.99\linewidth]{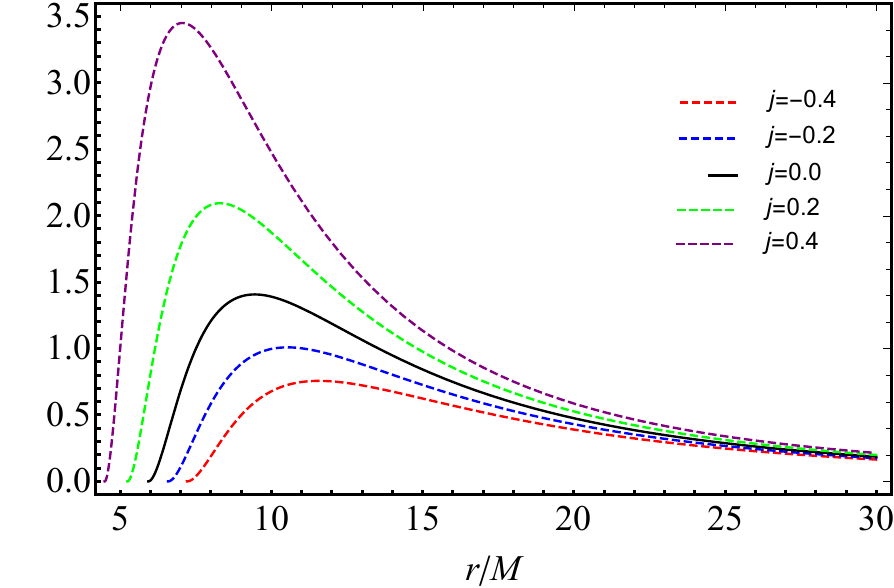}\\ }
\end{minipage}
\hfill
\begin{minipage}{0.50\linewidth}
\center{\includegraphics[width=0.95\linewidth]{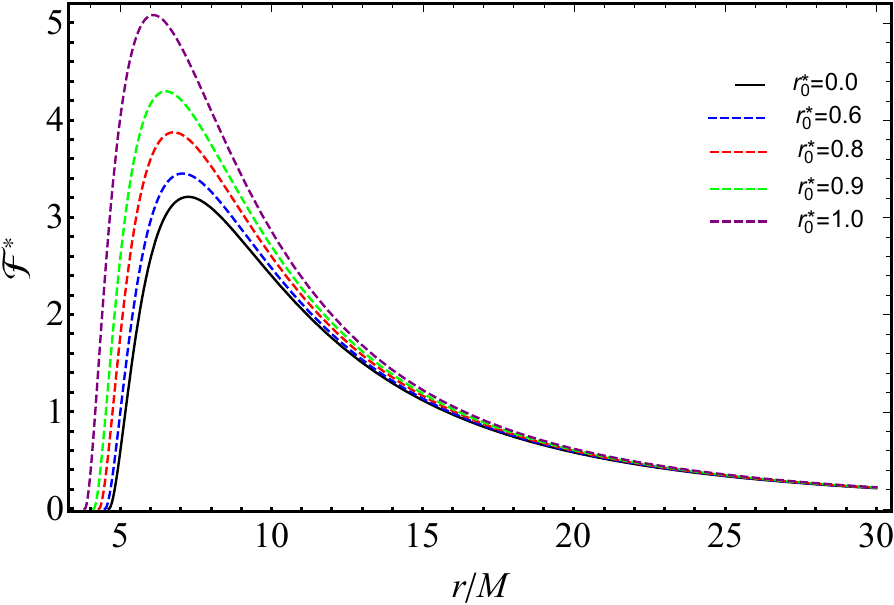}\\ }
\end{minipage}
\caption{Left panel: Radiative flux $\mathcal{F}^*$ multiplied by $10^{5}$ of the accretion disk versus normalized radial distance $r/M$ for rotating RBHs with $r_{0}^{*}=0.6$. Right panel: Radiative flux $\mathcal{F}^*$ multiplied by $10^{5}$ of the accretion disk versus normalized radial distance $r/M$ for RBHs with $j=0.4$.}
\label{fig:Flux0604}
\end{figure*}
\begin{figure*}[ht]
\begin{minipage}{0.49\linewidth}
\center{\includegraphics[width=0.98\linewidth]{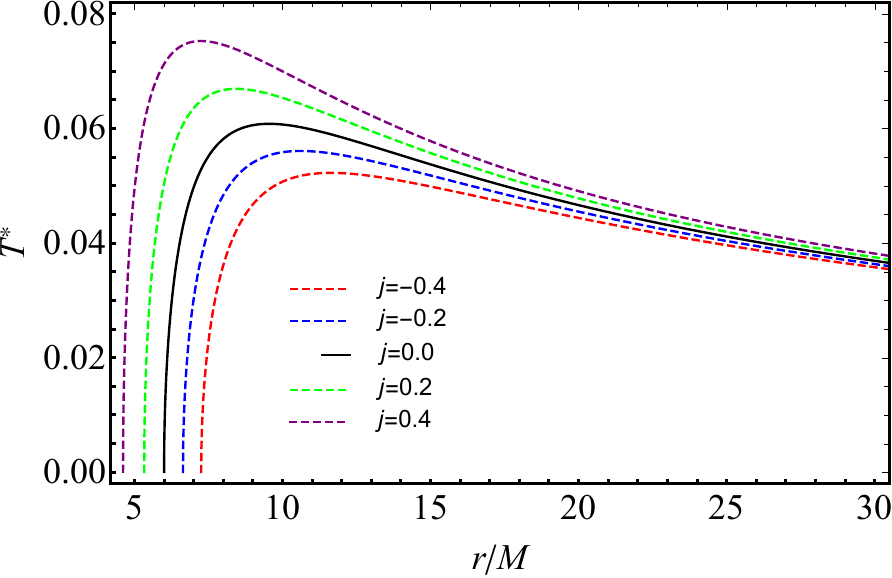}\\ }
\end{minipage}
\hfill
\begin{minipage}{0.50\linewidth}
\center{\includegraphics[width=0.98\linewidth]{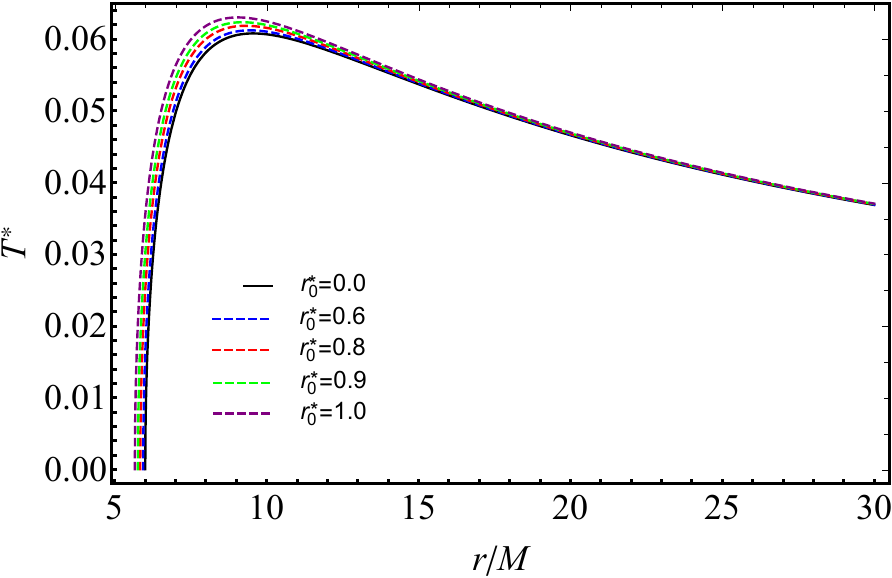}\\ }
\end{minipage}
\caption{Left panel: The temperature $T^{*}$ of accretion disks around the rotating RBHs with $r^*_{0}=0$. Right panel: The temperature $T^{*}$ of accretion disks around the RBHs with $j=0$.}
\label{fig:TKr0}
\end{figure*}

\begin{figure*}[ht]
\begin{minipage}{0.49\linewidth}
\center{\includegraphics[width=0.98\linewidth]{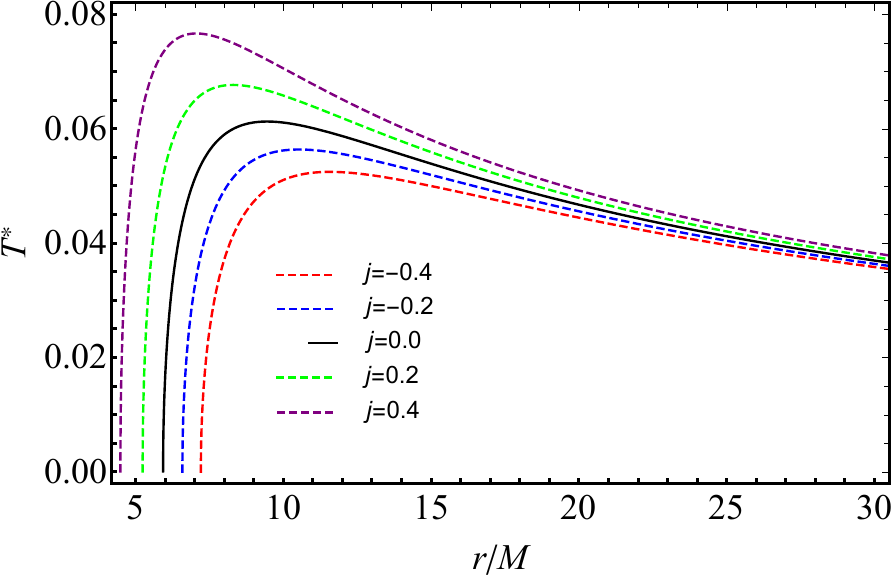}\\ }
\end{minipage}
\hfill
\begin{minipage}{0.50\linewidth}
\center{\includegraphics[width=0.98\linewidth]{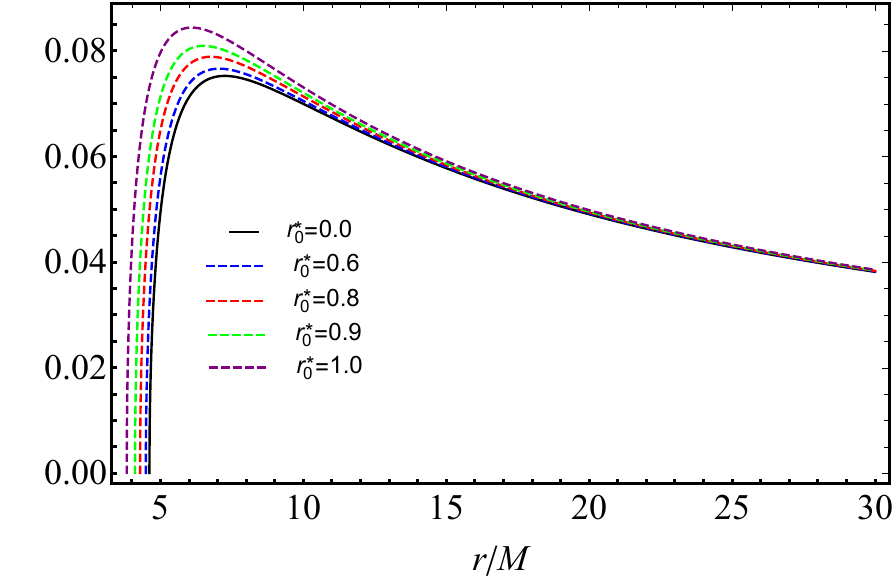}\\ }
\end{minipage}
\caption{Left panel: The temperature $T^{*}$ of accretion disks around the rotating RBHs with $r^*_{0}=0.6$. Right panel: The temperature $T^{*}$ of accretion disks around the RBHs with $j=0.4$.}
\label{fig:T0604}
\end{figure*}
\begin{figure*}[ht]
\begin{minipage}{0.49\linewidth}
\center{\includegraphics[width=0.98\linewidth]{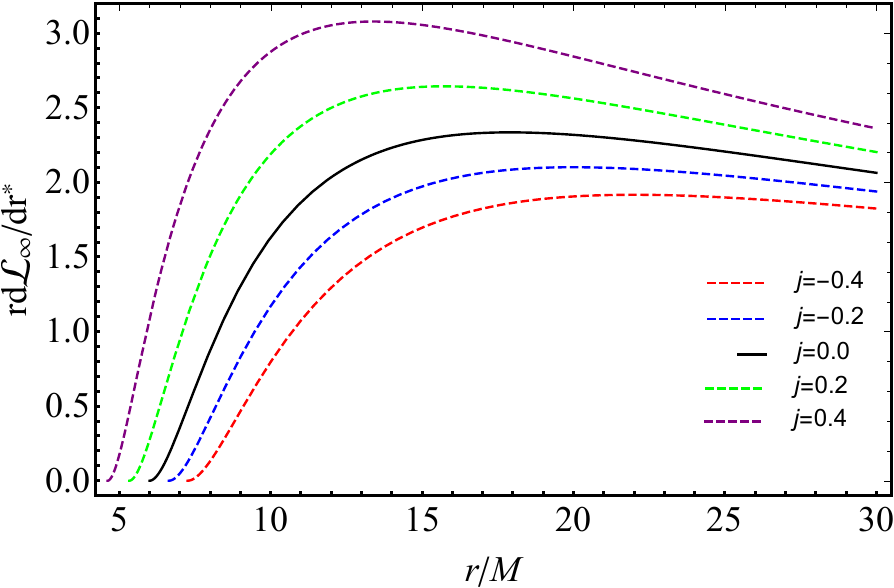}\\ }
\end{minipage}
\hfill
\begin{minipage}{0.50\linewidth}
\center{\includegraphics[width=0.98\linewidth]{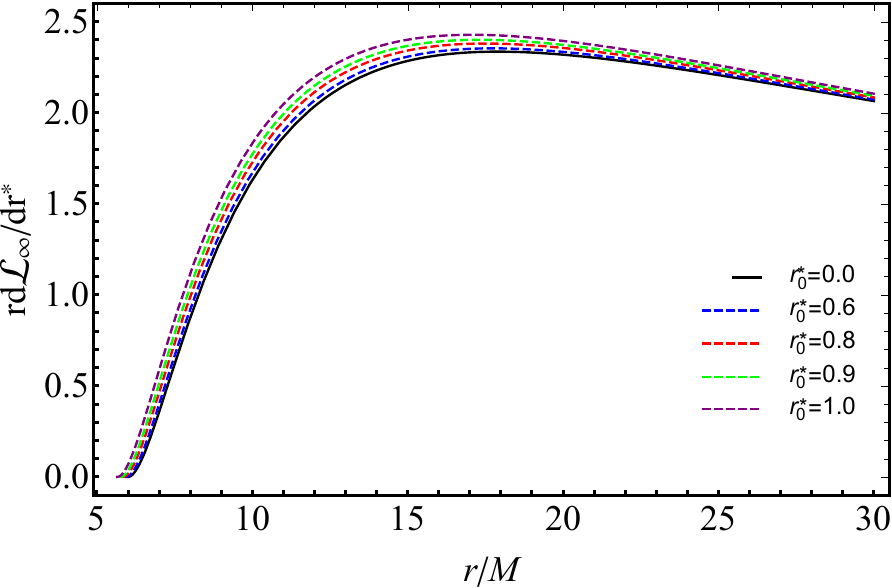}\\ }
\end{minipage}
\caption{Left panel: Differential luminosity multiplied by $10^{2}$ of the accretion disk versus radial distance $r$ normalized in units of total mass $M$ for rotating RBHs with $r_{0}^{*}=0$. Right panel: Differential luminosity multiplied by $10^{2}$ of the accretion disk versus radial distance $r$ normalized in units of total mass $M$ for RBHs with $j=0$.}
\label{fig:Dif_lumKr0}
\end{figure*}
\begin{figure*}[ht]
\begin{minipage}{0.49\linewidth}
\center{\includegraphics[width=0.98\linewidth]{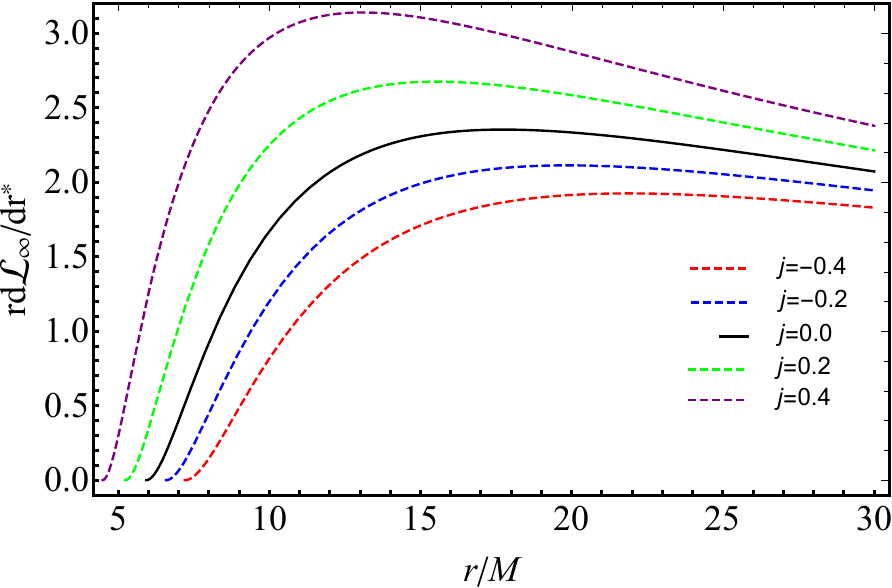}\\ }
\end{minipage}
\hfill
\begin{minipage}{0.50\linewidth}
\center{\includegraphics[width=0.98\linewidth]{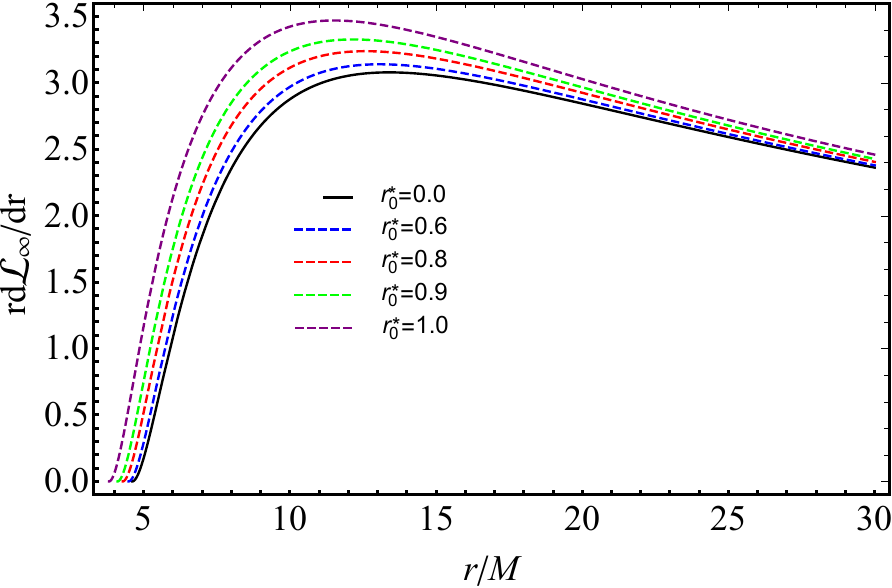}\\ }
\end{minipage}
\caption{Left panel: Differential luminosity multiplied by $10^{2}$ of the accretion disk versus radial distance $r$ normalized in units of total mass $M$ for rotating RBHs with $r_{0}^{*}=0.6$. Right panel: Differential luminosity multiplied by $10^{2}$ of the accretion disk versus radial distance $r$ normalized in units of total mass $M$ for RBHs with $j=0.4$.}
\label{fig:Dif_lum0604}
\end{figure*}

\begin{figure*}[ht]
\begin{minipage}{0.49\linewidth}
\center{\includegraphics[width=0.98\linewidth]{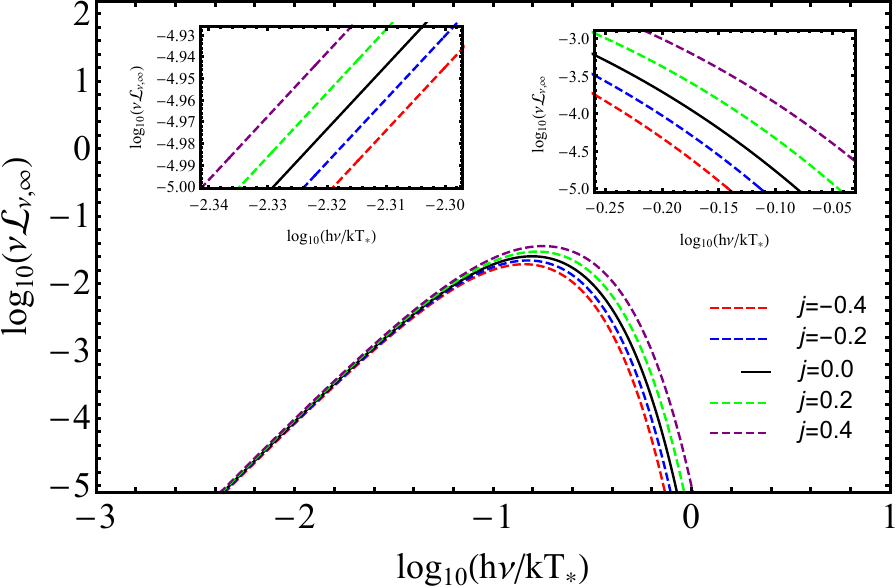}\\ }
\end{minipage}
\hfill
\begin{minipage}{0.50\linewidth}
\center{\includegraphics[width=0.98\linewidth]{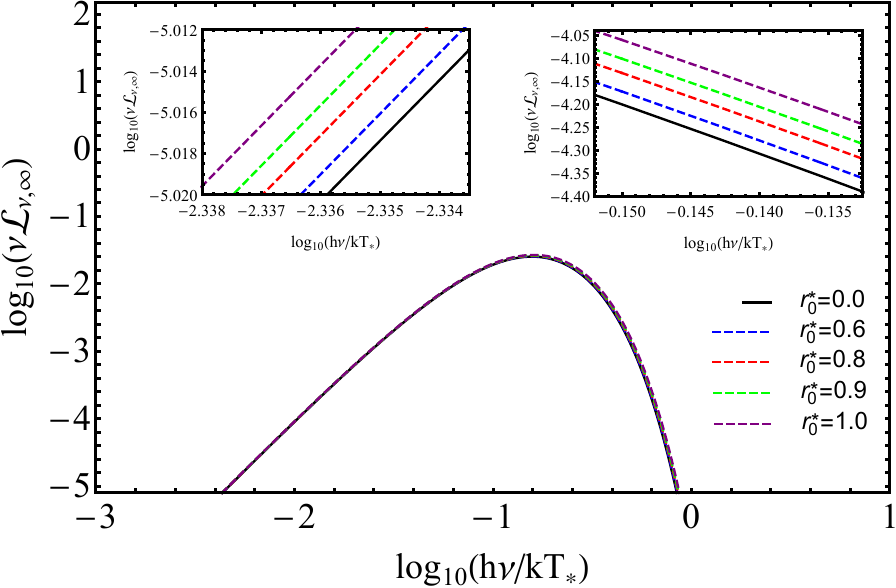}\\ }
\end{minipage}
\caption{Left panel: Spectral luminosity versus frequency of the emitted radiation for blackbody emission of the accretion disk for rotating RBHs with $r_{0}^{*}=0$.Right panel: Spectral luminosity versus frequency of the emitted radiation for blackbody emission of the accretion disk for RBHs with $j=0$.}
\label{fig:SpectralKr0}
\end{figure*}
\begin{figure*}[ht]
\begin{minipage}{0.49\linewidth}
\center{\includegraphics[width=0.98\linewidth]{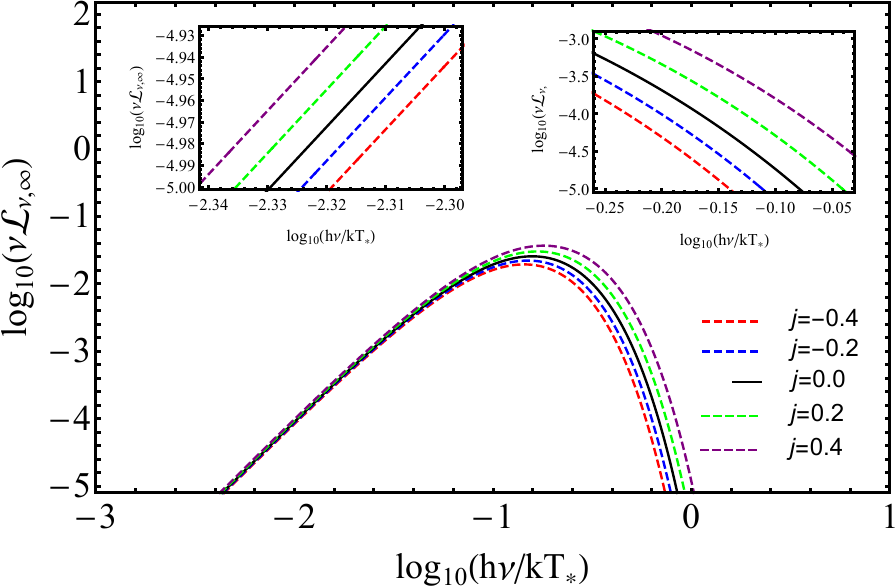}\\ }
\end{minipage}
\hfill
\begin{minipage}{0.50\linewidth}
\center{\includegraphics[width=0.98\linewidth]{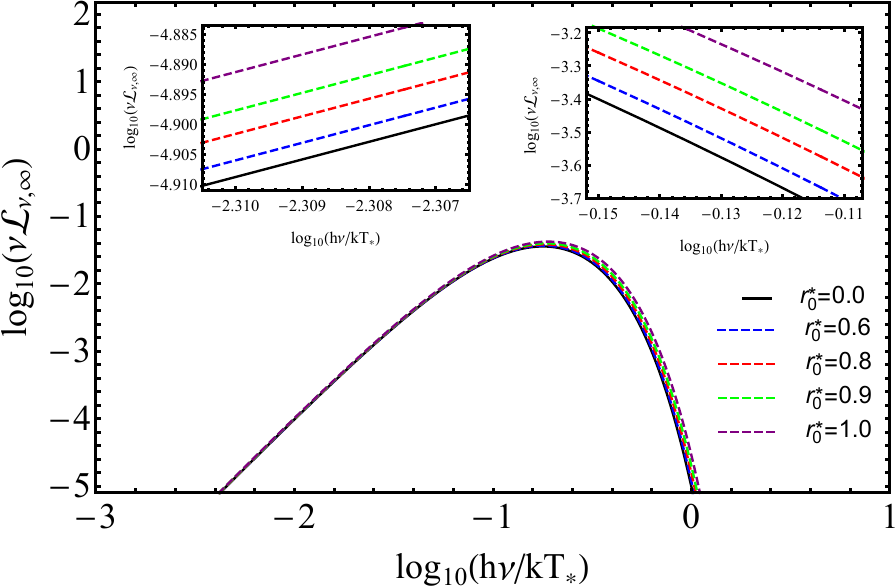}\\ }
\end{minipage}
\caption{Left panel: Spectral luminosity versus frequency of emitted radiation for black-body emission of the accretion disk for rotating RBHs with $r_{0}^{*}=0.6$. Right panel: Spectral luminosity versus frequency of emitted radiation for black-body emission of the accretion disk for RBHs with $j=0.4$.}
\label{fig:Spectral0604}
\end{figure*}

\subsection{Theoretical analysis}

In Fig.~\ref{fig:FluxKr0} we plotted the radiative flux as a function of the normalized radial coordinate $r/M$ for rotating RBHs with fixed $r_0^*=0$ and different values of $j$, which corresponds to the Kerr metric (left panel) and for static RBHs with fixed $j=0$ and different values of $r_{0}^{*}$, which corresponds to the Hayward metric (right panel).  In the left panel, for positive values of $j$ we have a larger flux in all the ranges of $r$ with respect to the Schwarzschild space-time, and for negative values of $j$ we have a smaller flux in all the ranges of $r$ than in the Schwarzschild solution. In the right panel the radiative flux in the Hayward space-time with positive values of $r_{0}^{*}$ is always larger than in the Schwarzschild spacetime in the entire range of $r$.

In Fig.~\ref{fig:Flux0604} we illustrate the radiative flux as a function of the normalized radial coordinate for rotating RBHs in analogy to Fig.~\ref{fig:FluxKr0} with the only exception, we fixed $r_{0}^{*} = 0.6$ and left $j$ to be arbitrary (left panel), while in the right panel, we fixed $j = 0.4$ and left $r_{0}^{*}$ to vary. As can be seen in the left panel, the rotating Hayward black holes with the same values of $j$, as in  Fig. ~\ref{fig:FluxKr0}, will have larger radiative flux with respect to the Kerr black hole. In the right panel, the radiative flux for rotating Hayward BHs with fixed value of $j=0.4$ and different values of $r_{0}^{*}$ is larger than static Hayward BHs (see the right panel of Fig.~\ref{fig:FluxKr0} ).

In Fig.~\ref{fig:TKr0} we plotted dimensionless temperature $T^{*}$ of accretion disks around the rotating RBHs with $r_{0}^{*}=0$, which as earlier, corresponds to the Kerr spacetime (left panel) and for RBHs with $j=0$ (right panel) which corresponds to the Hayward spacetime. In the left panel, the temperature for positive (negative) values of $j$ is larger (smaller) than in the Schwarzschild space-time (black solid curve). In the right panel, the temperature for all positive values $r_{0}^{*}$ is always larger than in the Schwarzschild space-time (black solid curve).

In Fig. ~\ref{fig:T0604} we present the dimensionless $T^{*}$ of accretion disks around RBHs in analogy to Fig.~\ref{fig:TKr0}, but with the only exception: here in the left panel we fix $r_{0}^{*}=0.6$ and vary $j$ and in the right panel we fix $j=0.4$ and vary $r_{0}^{*}$. Thus, the temperature in the accretion disk around rotating Hayward BHs is higher than that around Kerr BHs with identical $j<1$ (left panel). In the right panel, the flux of the disk around rotating Hayward BHs with fixed $j=0.4$ and different $r_{0}^{*}$ is higher than in static the Hayward metric with the same $r_{0}^{*}$.

In Fig.~\ref{fig:Dif_lumKr0} we present the differential luminosity as a function of the normalized radial coordinate for rotating RBHs, where we fix $r_{0}^{*}=0$ and vary $j$ (left panel) and fix $j=0$ and vary $r_{0}^{*}$ (right panel). Here, we observe a similar behavior of the differential luminosity in analogy to the radiative flux shown in Fig.~\ref{fig:FluxKr0}. It is due to the fact that both quantities are related by Eq.~\ref{eq:difflum}. Therefore, everything observed in the flux automatically translates into the differential luminosity. The same description is true for Fig.~\ref{fig:Dif_lum0604} which directly related to Fig. ~\ref{fig:Flux0604}.

In Fig.~\ref{fig:SpectralKr0}, we depict the spectral luminosity $\mathcal{L}_{\nu,\infty}$, defined by Eq.~\eqref{eq:speclum}, as a function of the frequency of radiation emitted by the accretion disk. The left panel shows rotating RBHs with fixed $r_{0}^{*}=0$ and arbitrary $j$, which corresponds to the Kerr metric, while the right panel represents RBHs with fixed $j=0$ and arbitrary $r_{0}^{*}$, which corresponds to the Hayward metric, as mentioned above. Thus, for co-rotating (counter-rotating) orbits the spectral luminosity of the accretion disk is larger (smaller) around the Kerr black hole with respect to the Schwarzschild one (left panel). The spectral luminosity of the disk around the static Hayward black hole with $r^*_0>0$  is always larger than around the Schwarzschild black hole. This fact is related to the value of $r_{ISCO}$. Unlike in the Schwarzschild spacetime, in the Hayward black hole with $r^*_0>0$ circular orbits can be closer to the central object.

Furthermore, in Fig.~\ref{fig:Spectral0604}, we see that the behavior of the spectral luminosity is similar to that presented in Fig. ~\ref{fig:SpectralKr0}. However, the inclusion of $r_0^*=0.6$ in the left panel and $j=0.4$ in the right panel increases the values of the spectral luminosity in both panels. Thus, we confirm that in the field of rotating Hayward black holes with $j<1$ and $r^*_0>0$ accretion disks can posses larger luminosity with respect to the Kerr metric.


\section{Final remarks}\label{sez5}

We here considered the spectral and  thermodynamic properties of accretion disks around rotating regular solutions. We explored these features based on the fact that accretion disk luminosity can be used to discriminate the type of spacetime that models a given BH or, more broadly, compact objects, BH mimickers, and so forth.

Consequently, we here investigated the effects that a generic  RBH solution, without the presence of cosmological constant, induces on the physics of accretion disk, comparing our findings with the corresponding singular picture that refers to the Kerr solution.

In such a way, we also highlighted the possibility of distinguishing rotating RBHs from the Kerr BH expectations. Our goal was to \emph{disentangle} regular from singular solution expected outcomes, showing the main differences in theoretical and measurable effects.

As a remarkable fact, we compared our outcomes even with the cases of non-rotating RBHs, showing the discrepancies that one expects to find comparing our solution with previous literature, i.e, with Hayward RBH.

To do so, we precisely worked out the accretion disk characteristics for rotating RBHs and, in particular, we computed the behaviors of neutral test particles within the circular geodesics. In this respect we evaluated the ISCO radius, radiative flux, differential luminosity and spectral luminosity.

Further, we estimated the efficiency of converting mass into energy of the accretion disks, emphasizing the precise departures from the Kerr solution.

As a landscape for accretion disk, we here invoked the simplest standard model of thin accretion disk, assuming the Novikov-Thorne-Page approach.

In particular, as a remarkable consequence of our outputs, we obtain that the luminosity of both the accretion disk and the efficiency turned out to be larger for RBHs if compared with Kerr metric predictions for fixed small values of the spin parameter $j<1$. Analogous results have been obtained on spectra and thermodynamic quantities. A clear difference than the Kerr metric and non-rotating RBH has been therefore emphasized.

In view of our findings, future developments will focus on alternative RBHs, involving additional physical parameters. Further, we will investigate alternatives to the Novikov-Thorne-Page scenario, including extra parameters that describe the accretion disk.

\begin{acknowledgments}
TK acknowledges the Grant No. AP19174979, YeK acknowledges Grant No. AP19575366 and KB and OL acknowledge Grant No. AP19680128 from the Science Committee of the Ministry of Science and Higher Education of the Republic of Kazakhstan. The authors are particularly thankful to Daniele Malafarina for his suggestions that significantly helped to improve the quality of this manuscript.
\end{acknowledgments}

\appendix

\section*{Kinematic properties of our solution}

For completeness we illustrate here the angular velocity $\Omega$, angular momentum $L$ and energy $E$ per unit mass of test particles on circular orbits in the RBH spacetime depending on the values of the angular momentum $j$ and parameter $r_0^*$.

\begin{figure*}[h!]
\begin{minipage}{0.49\linewidth}
\center{\includegraphics[width=0.99\linewidth]{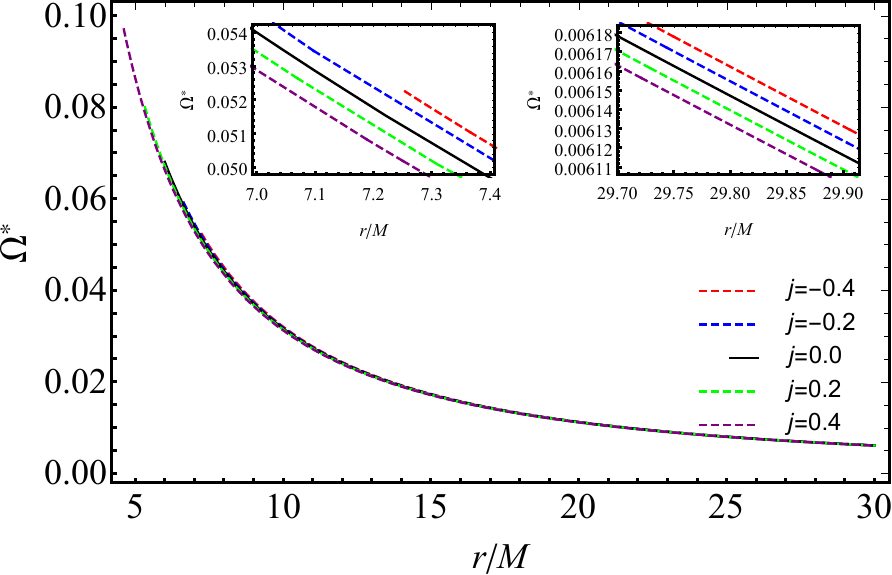}\\ }
\end{minipage}
\hfill
\begin{minipage}{0.50\linewidth}
\center{\includegraphics[width=0.95\linewidth]{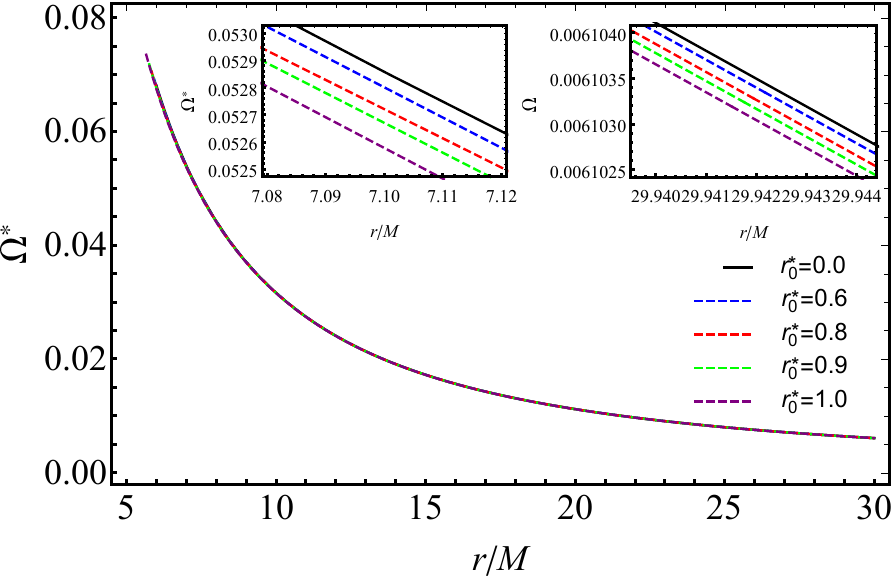}\\ }
\end{minipage}
\caption{Left panel: Angular velocity of test particles versus radial distance $r$ normalized in units of total mass $M$ for rotating regular holes with $r_{0}^{*}=0$. Right panel: Angular velocity of test particles versus radial distance $r$ normalized in units of total mass $M$ for regular black holes with $j=0$.}
\label{fig:omegaKr0}
\end{figure*}

In Fig.~\ref{fig:omegaKr0} we show the orbital angular velocity $\Omega^*(r)$ of test particles as a function of the normalized radial coordinate $r/M$ for Kerr with  different values of $j$=[-0.4, -0.2, 0, 0.2, 0.4] (left panel)  and  for static regular black holes with different values of $r_{0}^{*}$=[0, 0.6, 0.8, 0.9, 1] (right panel).

\begin{figure*}[h!]
\begin{minipage}{0.49\linewidth}
\center{\includegraphics[width=1\linewidth]{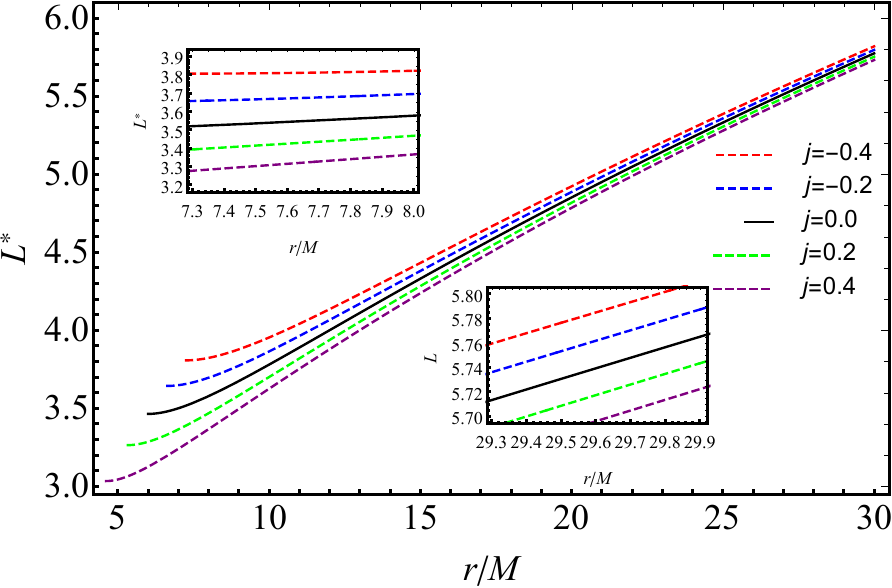}\\ }
\end{minipage}
\hfill
\begin{minipage}{0.50\linewidth}
\center{\includegraphics[width=0.95\linewidth]{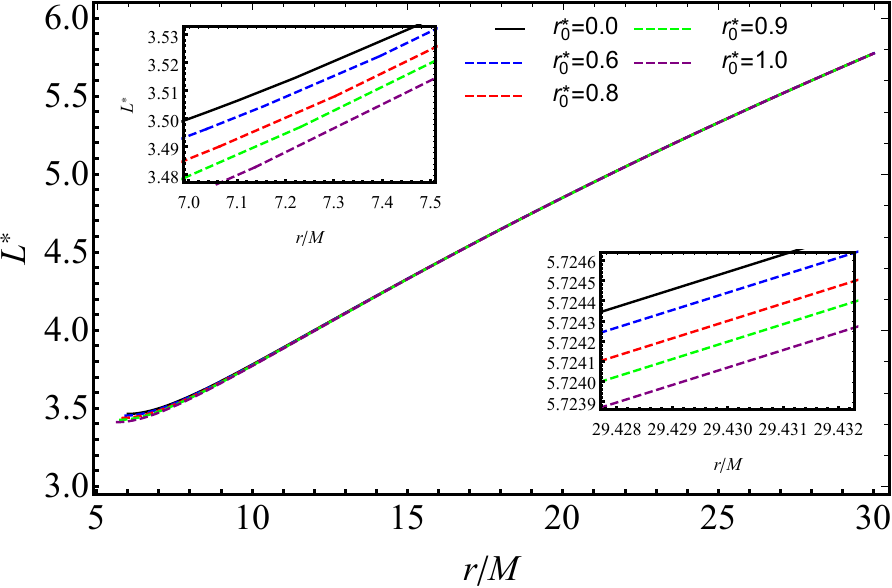}\\ }
\end{minipage}
\caption{Left panel: Angular momentum $L^*$ of test particles versus normalized radial distance $r/M$ for rotating regular black holes with $r_{0}^{*}=0$. Right panel: Angular momentum $L^*$ of test particles versus normalized radial distance $r/M$ for regular black holes with $j=0$.}
\label{fig:LKr0}
\end{figure*}

In Fig.~\ref{fig:LKr0} we constructed the dimensionless orbital angular momentum $L^*(r)$ of test particles as a function of the normalized radial coordinate $r/M$ for Kerr, i.e. $r_{0}^{*}=0$, with different values $j$ (left panel) and for static RBHs with different values of $r_{0}^{*}$ (right panel).

\begin{figure*}[h!]
\begin{minipage}{0.49\linewidth}
\center{\includegraphics[width=0.99\linewidth]{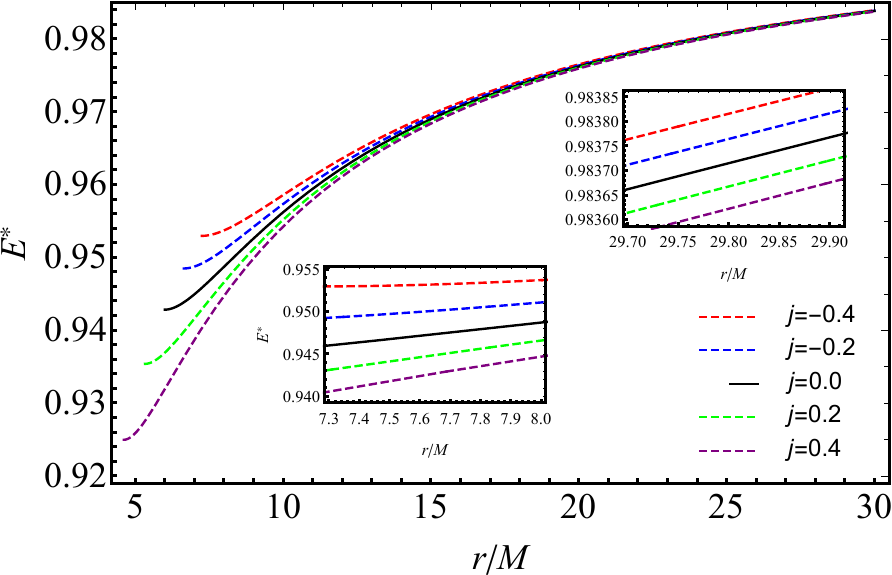}\\ }
\end{minipage}
\hfill
\begin{minipage}{0.50\linewidth}
\center{\includegraphics[width=0.95\linewidth]{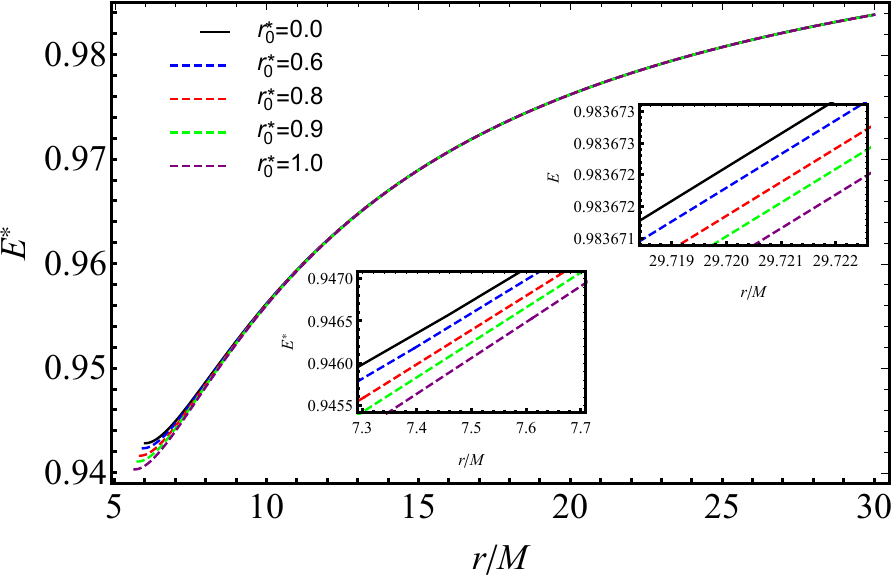}\\ }
\end{minipage}
\caption{Left panel: Energy $E^*$ of test particles versus normalized radial distance $r/M$ for rotating regular  black holes with $r_{0}^{*}=0$. Right panel: Energy $E^*$ of test particles versus normalized radial distance $r/M$ for regular black holes with $j=0$.}
\label{fig:EnergyKr0}
\end{figure*}

In Fig.~\ref{fig:EnergyKr0} we constructed the energy per unit mass $E^*$ of test particles as a function of the normalized radial coordinate $r/M$ for Kerr (left panel) and for static regular black holes (right panel).

\newpage

\end{document}